\documentclass[twocolumn,superscriptaddress,amssymb,aps, prc,nobibnotes,floatfix,showpacs,nofootinbib]{revtex4-2}
\usepackage{amsmath}
\usepackage{amssymb}
\usepackage{graphicx}
\usepackage{color}
\usepackage{float}
\usepackage{longtable}
\usepackage{afterpage}
\usepackage{mathptmx}
\usepackage[colorlinks=true, allcolors=blue]{hyperref}

\usepackage{epstopdf}
\usepackage{soul}

\begin{document}
	
\title{In-beam $\gamma$-spectroscopy of the transitional nucleus $^{\textbf{217}}$Ac}
\author{Dhananjaya Sahoo}
\affiliation{Department of Physics, Indian Institute of Technology Roorkee, Roorkee - 247667, INDIA}
\author{A. Y. Deo}
\email{\textcolor{black}{Corresponding author:} ajay.deo@ph.iitr.ac.in}
\affiliation{Department of Physics, Indian Institute of Technology Roorkee, Roorkee - 247667, INDIA}
\author{Madhu}
\affiliation{Department of Physics, Indian Institute of Technology Roorkee, Roorkee - 247667, INDIA}
\author{Khamosh Yadav}
\thanks{Present address: Department of Physics, Indian Institute of Technology Ropar, Rupnagar - 140001, INDIA}
\affiliation{Department of Physics, Indian Institute of Technology Roorkee, Roorkee - 247667, INDIA}
\author{S. S. Tiwary}
\thanks{Present address: Department of Physics, School of Basic Sciences,\\ Manipal University Jaipur, Jaipur - 303007, INDIA}
\affiliation{Department of Physics, Indian Institute of Technology Roorkee, Roorkee - 247667, INDIA}
\author{P. C. Srivastava}
\affiliation{Department of Physics, Indian Institute of Technology Roorkee, Roorkee - 247667, INDIA}
\author{R. Palit}
\affiliation{Department of Nuclear and Atomic Physics, 
	Tata Institute of Fundamental Research, Mumbai - 400005, INDIA}
\author{S. K. Tandel}
\affiliation{Department of Physics, School of Natural Sciences, Shiv Nadar Institute of Eminence, Gautam Buddha Nagar - 201314, INDIA}
\author{Anil Kumar}
\affiliation{Division of Nuclear Physics, Center for Computational Sciences, University of Tsukuba, Tsukuba - 3058577, JAPAN}
\author{P. Dey}
\affiliation{Department of Nuclear and Atomic Physics, 
	Tata Institute of Fundamental Research, Mumbai - 400005, INDIA}
\author{Biswajit Das}
\affiliation{Department of Nuclear and Atomic Physics, 
	Tata Institute of Fundamental Research, Mumbai - 400005, INDIA}
\author{Vishal Malik}
\affiliation{Department of Nuclear and Atomic Physics, 
	Tata Institute of Fundamental Research, Mumbai - 400005, INDIA}
\author{A. Kundu}
\affiliation{Department of Nuclear and Atomic Physics, 
	Tata Institute of Fundamental Research, Mumbai - 400005, INDIA}
\author{A. Sindhu}
\affiliation{Department of Nuclear and Atomic Physics, 
	Tata Institute of Fundamental Research, Mumbai - 400005, INDIA}
\author{S. V. Jadhav}
\affiliation{Department of Nuclear and Atomic Physics, 
	Tata Institute of Fundamental Research, Mumbai - 400005, INDIA}
\author{B. S. Naidu}
\affiliation{Department of Nuclear and Atomic Physics, 
	Tata Institute of Fundamental Research, Mumbai - 400005, INDIA}
\author{A. V. Thomas}
\affiliation{Department of Nuclear and Atomic Physics, 
	Tata Institute of Fundamental Research, Mumbai - 400005, INDIA}
\date{\hfill \today}
\begin{abstract}
 High-spin states in the transitional $^{217}$Ac nucleus are established up to 3.8 MeV excitation energy and $I^{\pi} = 41/2^{+}$ with the addition of around 20 new transitions. The structure of the yrast and near-yrast states below the $29/2^{+}$ isomer is revisited. The inconsistencies in the level schemes reported earlier are resolved. The level structure above the $29/2^{+}$ isomer is established for the first time. Large-basis shell-model calculations with the KHPE interaction are performed to compare the experimentally observed level energies with the theoretical predictions. A comparison with the systematics of the $N = 128$ isotones suggests that the yrast structures result from a weak coupling of the odd proton to the even-even $^{216}$Ra core, which is consistent with the shell-model configurations. Furthermore, $\alpha$-decay of the $29/2^{+}$ isomer is revisited and the decay scheme established from this work is discussed in the framework of the shell model.
\end{abstract}
\maketitle
\section{Introduction}
The low- and high-spin structures of atomic nuclei provide an ideal laboratory to study the effect of quantum many-body aspects. Various nuclear models based on quantum many-body effects aid in examining their applicability in the understanding of experimental observations in different regions of the nuclear chart \cite{Brown}. Spectroscopic studies of nuclei beyond the doubly-magic $^{208}$Pb reveal the existence of diverse structural phenomena. For instance, nuclei in vicinity of the shell closures have spherical or near-spherical shapes and their properties are interpreted in the framework of the shell model where nucleons occupy several single particle quantum states \cite{McGrory,Brown2,Blomqvist}. On the other hand, the addition of a sufficient number of nucleons beyond the shell closures distorts the spherical symmetry \cite{Ahmad}. Furthermore, the availability of the single particle orbitals with $\Delta j = \Delta l = 3$ i.e. ($ \pi i_{13/2}$ and $ \pi f_{7/2}$) and ($\nu g_{9/2}$ and $ \nu j_{15/2}$) and the interaction between them result in the long-range octupole-octupole correlations between the nucleons \cite{Ahmad,Butler,Butler2,Butler3}. Nuclei with $Z \approx 88$ and $N \approx 136$ show strong octupole correlations, leading to a reflection-asymmetric stable pear shape in such nuclei \cite{Gaffney,Pancholi}. The nuclei lying between the two extremes of shape are known as transitional nuclei. The transitional nuclei are of particular interest because they display features arising from the interplay between the two fundamental modes of excitation i.e. single particle and collective modes. Another intriguing feature of this region is the enhanced $\alpha$-decay widths for the ground and excited states. This indicates that the  $\alpha$-clustering effects play a significant role in the structure of the transitional nuclei \cite{Gai,Iachello,Daley1,Daley2,Toth,Debray}.

The high-spin states in the $N=128$ trans-lead isotones exhibit some interesting features \cite{Itoh, Dracoulis, Muralithar,Lonnroth,Lane,Garnsworthy,KYadav, Drigert}. Many high-spin states and isomers have been established in $^{214}$Rn \cite{Dracoulis, Lonnroth}, $^{215}$Fr \cite{KYadav, Drigert} and $^{216}$Ra \cite{Muralithar, Lonnroth, Itoh} and their level structures have been accounted well by the shell model. The availability of the high-spin orbitals for several valence protons and two neutrons outside the shell closures ($ Z = 82 $ and $ N = 126 $) and the effect of the p-n interaction give rise to the realization of isomeric states at high-spin \cite{Dracoulis}. In addition, the presence of high-spin isomers indicates the hindrance in the decay due to the change in single particle configurations and interaction of single-particle states with octupole-phonon vibrations \cite{Dracoulis, KYadav,Lane}. Some of these isomers decay via enhanced $E3$ transitions. The active orbitals which are involved in particle-octupole vibrational couplings also lead to the structural evolution towards octupole deformation in heavier nuclei \cite{Lane}. Furthermore, the low-spin yrast states in the nuclei with $Z \geq 82$ and $N \leq 126$ have contributions from the valence protons in the $1h_{9/2}, 1i_{13/2}$, and $2f_{7/2}$ orbitals. However, for high-spin yrast states, neutron core-excitations also play an important role. For example, in $^{210-212}$Rn \cite{Poletti1,Poletti2,Poletti3,Horn}, although neutron excitations across the shell closure at $N =126$ are energetically very expensive, the presence of high-j neutron orbitals and the compensating energy gain from the attractive p-n interaction allow such core excitations \cite{Dracoulis}. On the other hand, for the $N=128$ isotones, the contributions from core excitations vanish for low spin-states. However, core excitation may significantly contribute to the high-spin states $ \sim 33 \hbar$ in $^{216}$Ra \cite{Muralithar, Lonnroth}. 

The light Ac-isotopes in the $A \sim 220$ region were studied earlier for the spectroscopic investigations \cite{Debray, Cristancho, Schulz2, Debray2, Aiche}. Experimental signatures of octupole correlations i.e. alternating-parity sequences interlinked by enhanced $E1$ transitions and parity-doublet structures, were reported in $^{218-221}$Ac \cite{Debray,Cristancho,Schulz2,Aiche}. As one moves away from the octupole deformed region beyond the $A = 220$ towards the neutron shell closure at $N = 126$, the collective phenomena which dominate at higher neutron numbers, start disappearing.  For example, the level structures in $^{215}$Ac ($N = 126$) have been interpreted in terms of spherical shell model taking the single particle energies and the proton pairing forces into account \cite{Decman215Ac}. Thus, there exists a narrow region of nuclei around the $N \sim 130$ which links the two different regions of structural phenomena. Moreover, similar structural properties have also been noticed in the transitional francium isotopes  \cite{Byrne1,Byrne2,Debray3,KYadav,Pragati,Aiche2,Debray4,Liang}.  There is a striking structural resemblance between the low-spin states of $^{215}$Fr and $^{217}$Ac. Recently, Yadav \textit{et al.}, extended the level scheme of $^{215}$Fr to $55/2\hbar$ and the overall level structures were interpreted in the framework of spherical shell model \cite{KYadav}. On the other hand, the level scheme of $^{217}$Ac was earlier studied up to $29/2\hbar$ and the low-lying yrast structures were described in terms of spherical shell model \cite{Decman}. Therefore, the high-spin states of $^{217}$Ac need to be explored with a detailed spectroscopic investigation in order to understand the evolution of structure from spherical to well-deformed shape in the actinium isotopic chain. 
\section{EXPERIMENTAL DETAILS AND ANALYSIS PROCEDURE}
\label{section2}
High-spin states in $^{217}$Ac were studied via the $^{209}$Bi($^{12}$C,4n)$^{217}$Ac reaction. The $^{12}$C beam of 72 MeV  was delivered by the 14 UD Pelletron-LINAC Facility at Tata Institute of Fundamental Research, Mumbai. The target consisted of a self-supporting $^{209}$Bi foil of $\sim$ 4 mg/cm$^{2}$ thickness. The $\gamma$ rays produced in the de-excitation of the residual nuclei were detected using the Indian National Gamma Array (INGA). The array comprised of 18 Compton suppressed HPGe clover detectors positioned in seven rings at various angles with respect to the beam direction. In the present configuration, the detectors were distributed at different angles; with two at 65$^{\circ}$, three each at 40$^{\circ}$, 115$^{\circ}$, 140$^{\circ}$, 157$^{\circ}$, and four at 90$^{\circ}$. The time stamped data were acquired with two and higher-fold coincidence condition using 12-bit 100 MHz PIXIE-16 digital data acquisition system developed by XIA-LLC, USA \cite{PIXIE DAQ}. The energy calibration and relative efficiency measurements of the detection system were carried out using the standard $^{152}$Eu and $^{133}$Ba radioactive sources.  The calibrated data were written into a ROOT Tree format using a code developed at IIT Roorkee, which is based on the code MultipARameter time-stamped based COincidence Search (MARCOS) \cite{PIXIE DAQ,CERN-ROOT}. The ROOT Trees format were used to generate various two and three dimensional histograms for further analysis using RADWARE and ROOT \cite{RADWARE1, RADWARE2}.

In order to search for the new $\gamma$ rays, two- and higher-fold coincidence events, in which the coincident $\gamma$ rays were detected within 100 ns of each other, were used to generate a prompt symmetric $\gamma$-$\gamma$ matrix and a $\gamma$-$\gamma$-$\gamma$ cube. Furthermore, an asymmetric \textit{early}-\textit{delayed} $\gamma$-$\gamma$ matrix was constructed for the $\gamma$ rays detected within the 200-1200 ns with respect to each other. In such a matrix, the \textit{early} and \textit{delayed} $\gamma$ rays are decided on the basis of their relative electronic timings. This matrix was used to establish the correlations among the $\gamma$ rays across the isomeric states. The spectroscopic information of $^{217}$Ac was extracted on the basis of coincidence relationships among the observed $\gamma$ rays, their relative intensities, and multipolarities.

The multipolarities of the $\gamma$ rays were determined in order to ascertain the spin and parities of the excited nuclear states of the level scheme. The multipole order of a transition was extracted using the directional correlation from oriented states (DCO) ratio method based on the angular correlation analysis \cite{DCO}. For the DCO ratio measurements, an asymmetric $\gamma$-$\gamma$ matrix was constructed with the $\gamma$ rays detected at 40$^{\circ}$ and 140$^{\circ}$ ($\theta_{1}$) detectors on the X-axis and the coincident $\gamma$ rays at 90$^{\circ}$ ($\theta_{2}$) detectors on the Y-axis. The DCO ratio for the transition of interest ($\gamma_{1}$) can be defined as,
\begin{equation}	
	R_{DCO}\;(\gamma_{1}) = \frac{I(\gamma_{1}) \;at \; \theta_{1} \; gated \; by \; \gamma_{2} \; at \; \theta_{2} }{I(\gamma_{1}) \; at \; \theta_{2}\; gated \; by \; \gamma_{2} \; at \; \theta_{1}}
\end{equation}
where $I(\gamma_{1})$ denotes the intensity of $\gamma_{1}$ measured in coincidence with another transition ($\gamma_{2}$) of a known multipolarity. The $R_{DCO}$ value for a quadrupole(dipole) transition was found to be nearly 1.0(1.0) in stretched quadrupole(dipole) gating transitions whereas it was 2.0(0.5) in stretched dipole(quadrupole) gating transitions. The intermediate values of the DCO ratio indicate a mixed nature of the transition.

\begin{figure}[t]
	\begin{center}	
		\includegraphics[width=1\linewidth]{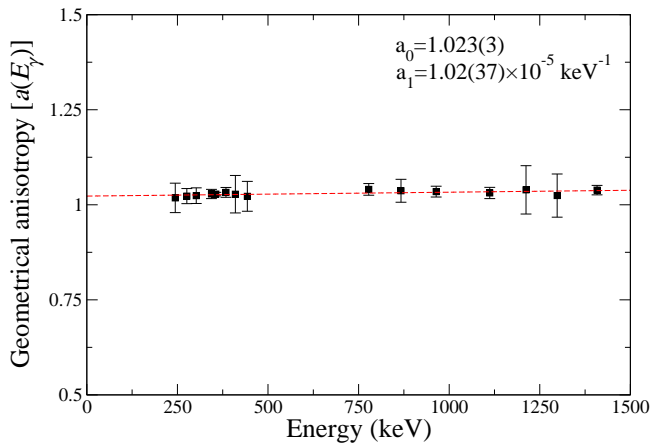}	
		\caption{The geometrical anisotropy factor $ a(E_{\gamma})$ as a function of $\gamma$-ray energies from $^{152}$Eu and $^{133}$Ba. The dotted line corresponds to the least-square fit to the data.}
		\label{fig1}
	\end{center}	
\end{figure}

The qualitative information about the electric or magnetic nature of a transition can be obtained from the linear-polarization measurements. The linear-polarization measurement depends on two factors: polarization asymmetry ($\Delta_{asym}$) \cite{PDCO1,PDCO2}, and polarization sensitivity [Q(E$_{\gamma}$)] \cite{PDCO2,PDCO3}. The polarization asymmetry for a transition of interests was deduced using the relation,
\begin{equation}
	\Delta_{asym} = \frac{a(E_{\gamma})N_{\perp}-N_{||}}{a(E_{\gamma})N_{\perp}+N_{||}}
\end{equation} 
where, $N_{||}(N_{\perp})$ represents counts of the Compton-scattered $\gamma$ rays detected in parallel(perpendicular) crystals of the $90^{\circ}$ clover detectors with respect to reaction plane. The parameter, $a(E_{\gamma})$, corresponds to the geometrical anisotropy; a correction due to the asymmetry in the response of the clover segments. The geometrical anisotropy, defined by $ a(E_{\gamma}) = \frac{N_{||}}{N_{\perp}}$, was determined as a function of energy for the unpolarized $\gamma$ rays from the $^{152}$Eu and $^{133}$Ba radioactive sources. In the present experiment, the least-square fit to the observed values of $ a(E_{\gamma})$ at different energies  $a(E_{\gamma}) = a_{0}+a_{1}E_{\gamma}$ yielded $a_{0}=1.023(3) $ and $a_{1}=1.02(37)\times10^{-5}$ keV$^{-1}$ as depicted in Fig. ~\ref{fig1}. 

\begin{figure}[b]
	\begin{center} 		 		
		\includegraphics[width=1\linewidth]{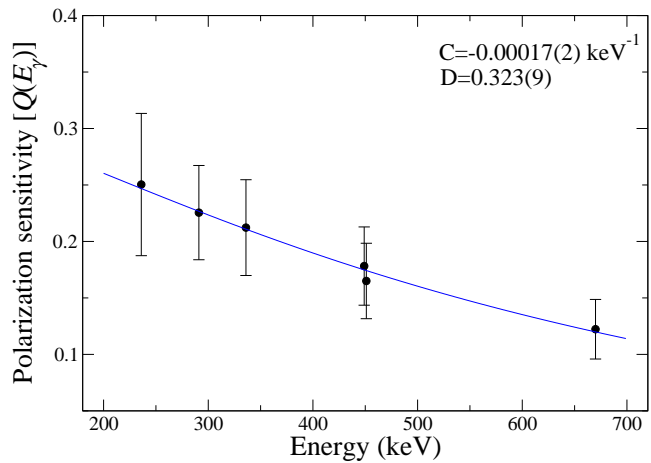}	
		\caption{Variation of polarization sensitivity as a function of $\gamma$-ray energies. The solid line depicts the least-square fit to the data.}
		\label{fig2}
	\end{center}
\end{figure}

\begin{figure*}[htb]
	\centering
	\includegraphics[width=1\linewidth]{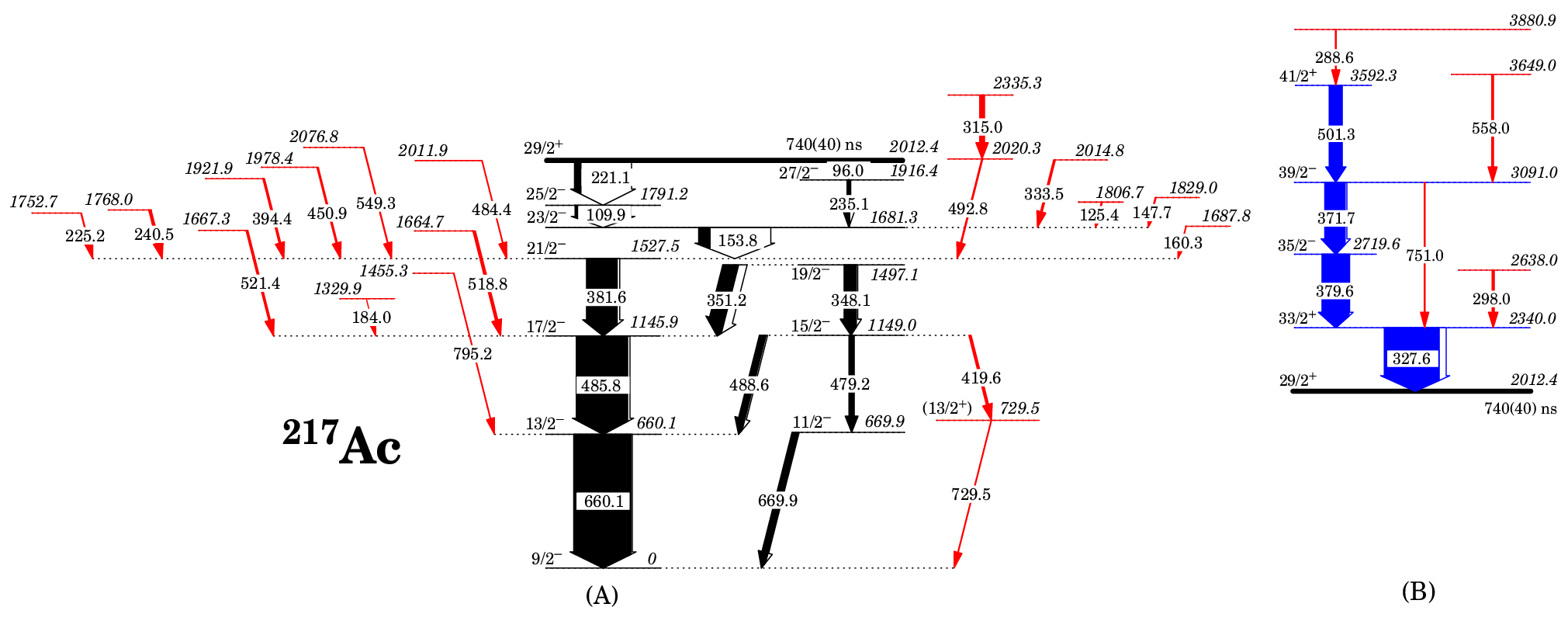}
	\caption{Partial level scheme of $^{217}$Ac illustrating the transitions (A) below and (B) above the 29/2$^{+}$ isomer. The widths of the closed and open areas of the arrows correspond to the relative intensity of the $\gamma$ rays and conversion electrons, respectively. The newly identified transitions are depicted in red color. The transitions whose placements were uncertain or inconsistent in earlier studies and have been confirmed in the present work are marked in blue color. The intensities of the transitions above the $29/2^{+}$ state are normalized with respect to that of the 328-keV transition, while those below are normalized with respect to the 660-keV transition. The multipolarity of the transitions in the part \textit{A} and the half life of the $29/2^{+}$ isomer are adopted from Ref. \cite{Decman}.}
	\label{fig3}
\end{figure*}

In order to determine the polarization asymmetry of the $\gamma$ rays of interest, two asymmetric matrices were generated using the $\gamma$ rays scattered either in parallel or perpendicular direction, with respect to the emission plane, of the $90^{\circ}$ clover detectors on one axis while the $\gamma$ ray energies recorded by all the other detectors along the other axis. It is understood that the polarization asymmetry is a function of Compton scattering probability as stated in Eq. (2) above. Furthermore, the probability of Compton scattering is itself energy dependent (i.e. function of $\gamma$ ray energy). Therefore, the energy dependence of polarization asymmetry ($\Delta_{asym}$) can be eliminated by normalizing it with the energy dependent polarization sensitivity [Q(E$_{\gamma}$)] \cite{SChatterjee} and thus defining the linear polarization as 
\begin{equation}
	P = \frac{\Delta_{asym}}{Q(E_{\gamma})} 
\end{equation}

The polarization sensitivity information can be extracted from the transitions of known multipolarity. The stretched \textit{E}2 transitions from $^{215}$Fr nucleus, one of the reaction products, were used for the present study. The sensitivity calculation for such $\gamma$ rays follows from the determination of the experimental polarization asymmetry ($\Delta_{asym}$) using Eq. (2) and theoretical linear polarization (\textit{P}), using the Klein-Nishina formula \cite{Klein-Nishina}, in terms of the angular distribution coefficients as described in the Ref. \cite{PDCO3, KYadav}. The experimental values of angular distribution coefficients for these transitions were adopted from the Ref. \cite{Drigert}. The variation of polarization sensitivity with $\gamma$-ray energy is illustrated in Fig. ~\ref{fig2} for the present data. A least-square fit to the observed sensitivity was performed using 
\begin{equation}
	Q(E_{\gamma}) = Q_{0}(E_{\gamma})(CE_{\gamma}+D)
\end{equation}
where $Q_{0}(E_{\gamma})$ represents the polarization sensitivity for a point scatterer and analyzer which is expressed as 
\begin{equation}
	Q_{0}(E_{\gamma}) = \frac{\alpha+1}{\alpha^{2}+\alpha+1}
\end{equation}
with $\alpha = E_{\gamma}/m_{e}c^{2}$. Here, $E_{\gamma}$ denotes the $\gamma$ ray energy and $m_{e}c^{2}$ is the rest mass energy of the electron. The values of corresponding parameters are C = -0.00017(2) keV$^{-1}$ and D = 0.323(9). A positive value of linear polarization indicates the electric nature of the transition of interest whereas the negative value corresponds to the magnetic nature. The mixed nature of a given transition may be inferred from a near-zero value of the linear polarization.
\section{results} \label{section3}
 In earlier studies, excited states in $^{217}$Ac were investigated using various heavy-ion fusion-evaporation reactions \cite{Decman,Gono}. For example, $^{205}$Tl($^{16}$O,4n), $^{206}$Pb($^{14}$N,4n) and $^{209}$Bi($^{12}$C,4n) reactions were employed by Decman \textit{et al.} for in-beam \textit{g}-factor measurements and coincidence studies between $\alpha$-, $\gamma$-, and conversion electron. In addition, pulsed beams of different pulse widths and repetition time were also used in these measurements. In their work, a total of thirteen $\gamma$ rays from eleven excited states up to an isomeric $I^{\pi}$ = 29/2$^{+}$ state with $T_{1/2}$ = 740(40) ns at 2013 keV were established. The \textit{g}-factor measurements were also performed to understand the structure of the isomeric state \cite{Decman}. In another study, high-spin states up to 3.3 MeV excitation energy and $I^{\pi}$ = 31/2$^{(+)}$ were studied by Gono \textit{et al.} using in-beam $\gamma$-ray spectroscopy and $\alpha-\gamma$ coincidences \cite{Gono}. However, the report by Gono \textit{et al.} do not provide any evidence for the 29/2$^{+}$ state.  Furthermore, alpha decay of the ground- and excited states including the isomer were also reported in $^{217}$Ac using in-beam studies \cite{Decman, Gono, Nomura}.
 
 In the present work, we report on the further extension of the level structure in $^{217}$Ac using in-beam $\gamma$-ray spectroscopic techniques. Figure~\ref{fig3} illustrates the partial level scheme of $^{217}$Ac resulting from the present analysis. The level scheme is extended up to 3.8 MeV excitation energy and $I^{\pi}$ = 41/2$^{+}$ with the addition of around 20 new transitions. The placements of the newly identified transitions in the level scheme is ascertained on the basis of coincidence relationships and intensity considerations. Moreover, the level structures below the 29/2$^{+}$ isomer have been revisited and are observed to be consistent with that reported in Ref. \cite{Decman}. Furthermore, some of the observations by Gono \textit{et al.} are found to be at variance with our work as well as with the one by Decman \textit{et al.} \cite{Gono,Akovali,Kondev,Decman}. Some of the inconsistencies below the 21/2$^{-}$ state were resolved earlier in the work by Decman \textit{et al.} \cite{Decman}. The spin and parity of the states, wherever possible, are determined on the basis of R$_{DCO}$ and linear-polarization measurements in the present work. Table ~\ref{tab:I} and ~\ref{tab:II} present the list of $\gamma$-ray energies, level energies, relative $\gamma$-ray intensities, and the corresponding multipolarities of the transitions below and above the isomeric state, respectively. It may be noted that the multipolarities listed in Table ~\ref{tab:I} are adopted from the Ref. \cite{Decman} while those reported in Table ~\ref{tab:II} are new and based on the measured values of R$_{DCO}$ and linear polarization. Apart from this, the $\alpha$-decay of the 29/2$^{+}$ isomer is revisited and it is found to be consistent with that reported in the Ref. \cite{Decman}. The detailed results are discussed below.
\subsection{Level structures}
\begin{figure}[b]
	\begin{center}
		\includegraphics[width=1\linewidth]{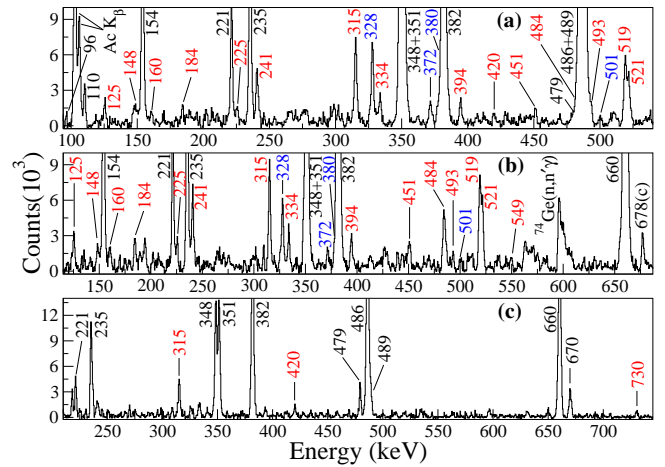}
		\caption{Coincidence $\gamma$-ray spectra depicting the transitions in the gate of the (a) 660-, (b) 486- and (c) 154-keV transitions. The new transitions are marked with red color. The transitions which were tentative in earlier studies and confirmed in the present work are shown in blue color. The contaminant transition labeled with (c) is from fission product $^{118}$Cd.}
		\label{fig4}
	\end{center}
\end{figure}

The level scheme of $^{217}$Ac is presented in two parts as shown in Fig.~\ref{fig3}. The part \textit{A} illustrates the states below the  29/2$^{+}$ isomer while the states above the isomer are depicted in the part \textit{B}. The two cascades of the $E2$ transitions interconnected via the mixed $M1+E2$ transitions were observed above the ground state as shown in the part \textit{A} of the level scheme \cite{Decman}. It was noted that the 660-keV transition was the most intense transition among all the transitions that fed the ground state $I^{\pi}$ = 9/2$^{-}$. This was found to be consistent with that reported in the Ref. \cite{Decman}. Figure \ref{fig4} represents $\gamma$ rays in coincidence with the 660-, 486- and 154-keV transitions. The present analysis confirms all the earlier placements below the 29/2$^{+}$ isomer \cite{Decman}. The spin-parity of the ground state and the excited states (see Table ~\ref{tab:I}) below the 29/2$^{+}$ isomer were adopted from the previous study by Decman \textit{et al.} \cite{Decman}.

\begin{table*}[t]
	\caption{\label{tab:I}Table of $\gamma$-ray energies ($E_{\gamma}$), level energies ($E_{i}$), spin-parity of the initial ($I^{\pi}_{i}$) and final ($I^{\pi}_{f}$) states, relative $\gamma$-ray intensities ($I_{\gamma}$), and  multipolarities  of the transitions below the 29/2$^{+}$ isomer in $^{217}$Ac. The relative $\gamma$-ray intensities are normalized with respect to that of the 660-keV transition. The listed spin-parities and multipolarities are adopted from the the Ref. \cite{Decman} except for the state  at 730 keV for which the $I^{\pi}$ assignment is based on the systematics. Both statistical and systematic uncertainties have been included in the uncertainty associated with the $\gamma$-ray energies and relative $\gamma$-ray intensities. The systematic uncertainty in $I_{\gamma}$ is considered to be 5$\%$ of $I_{\gamma}$.}
	
	\begin{ruledtabular}
		\begin{tabular}{cccccc}
			$E_{\gamma}$(keV)  &  $E_{i}$(keV)  &  $I^{\pi}_{i}$   &   $I^{\pi}_{f}$   &   $I_{\gamma}$   &  Multipolarity \\			
			\hline		
			96.0(3) & 2012.4(3) & 29/2$^{+}$ & 27/2$^{-}$ &  2.4(3) & $M1+E2$ \\
			109.9(2) & 1791.2(2) & 25/2$^{-}$ & 23/2$^{-}$ &  6.6(6) & $M1$ \\
			125.4(3) & 1806.7(3) &  & 23/2$^{-}$ & 1.6(2) & \\
			147.7(3) & 1829.0(3) &  & 23/2$^{-}$ & 1.5(2) & \\
			153.8(1) & 1681.3(1) & 23/2$^{-}$ & 21/2$^{-}$ & 20.0(13) & $M1+E2$ \\		
			160.3(4) & 1687.8(4) &  & 21/2$^{-}$ & 0.7(1) & \\
			184.0(5) & 1329.9(5) &  & 17/2$^{-}$ & 0.4(1) & \\
			221.1(2) & 2012.4(3) & 29/2$^{+}$ & 25/2$^{-}$ & 11.3(9) & $M2$ \\
			225.2(3) & 1752.7(3) &  & 21/2$^{-}$ & 1.2(1) & \\
			235.1(3) & 1916.4(3) & 27/2$^{-}$ & 23/2$^{-}$ & 4.9(5) & $E2$ \\
			240.5(3) & 1768.0(3) &  & 21/2$^{-}$ & 4.1(3) & \\
			315.0(2) & 2335.3(2) &  &  & 8.3(5) & \\
			333.5(3) & 2014.8(3) &  & 23/2$^{-}$ & 2.9(2) & \\
			348.1(1) & 1497.1(1) & 19/2$^{-}$ & 15/2$^{-}$ & 21.5(17) & $E2$ \\
			351.2(1) & 1497.1(1) & 19/2$^{-}$ & 17/2$^{-}$ & 26.9(20) & $M1+E2$ \\
			381.6(1) & 1527.5(1) & 21/2$^{-}$ & 17/2$^{-}$ & 53.6(32) & $E2$ \\
			394.4(3) & 1921.9(3) &  & 21/2$^{-}$ & 2.8(3) & \\
			419.6(3) & 1149.0(5) & 15/2$^{-}$ & (13/2$^{+}$)\footnotemark[1] & 3.8(5) & \\
			450.9(5) & 1978.4(5) &   & 21/2$^{-}$ & 2.4(2) & \\
			479.2(2) & 1149.0(5) & 15/2$^{-}$ & 11/2$^{-}$ & 8.7(9) & $E2$ \\
			484.4(5) & 2011.9(5) &  & 21/2$^{-}$ &   & \\
			485.8(1) & 1145.9(1) & 17/2$^{-}$ & 13/2$^{-}$ & 89.9(51) & $E2$ \\
			488.6(2) & 1149.0(5) & 15/2$^{-}$ & 13/2$^{-}$ & 11.4(12) & $M1+E2$ \\
			492.8(3) & 2020.3(3) &  & 21/2$^{-}$ & 1.8(3) & \\
			518.8(3) & 1664.7(3) &  & 17/2$^{-}$ & 4.2(3) & \\
			521.4(3) & 1667.3(3) &  & 17/2$^{-}$ & 3.1(2) & \\
			549.3(5) & 2076.8(5) &  & 21/2$^{-}$ & 1.2(2) & \\
			660.1(1) & 660.1(1) & 13/2$^{-}$ & 9/2$^{-}$ & 100.0(57) & $E2$ \\
			669.9(2) & 669.9(2) & 11/2$^{-}$ & 9/2$^{-}$ & 11.1(11) & $M1+E2$ \\ 
			729.5(4) & 729.5(4) & (13/2$^{+}$)\footnotemark[1] & 9/2$^{-}$ & 1.0(2)\footnotemark[2] & ($M2)$\footnotemark[1] \\
			795.2(5) & 1455.3(5) &  & 13/2$^{-}$ & 0.8(1) & \\	
		\end{tabular}
	\end{ruledtabular}
	\footnotetext[1]{Based on systematics.}\label{a}
	\footnotetext[2]{From intensity balance.}
\end{table*}

\begin{table*}[htb]
	\caption{\label{tab:II} Table of $\gamma$-ray energies ($E_{\gamma}$), level energies ($E_{i}$), spin-parity of the initial ($I^{\pi}_{i}$) and final ($I^{\pi}_{f}$) states, relative $\gamma$-ray intensities ($I_{\gamma}$), DCO ratios ($R_{DCO}$), linear polarization ($P$) and deduced multipolarities  of the transitions above the 29/2$^{+}$ isomer in $^{217}$Ac. The $R_{DCO}$ values of the transitions are determined in both dipole (D) and quadrupole (Q) gate. The relative $\gamma$-ray intensities are normalized with respect to the intensity of 328-keV transition. Both statistical and systematic uncertainties are included in the uncertainty associated with the $\gamma$-ray energies, relative $\gamma$-ray intensities and $R_{DCO}$ values. The systematic uncertainty in $I_{\gamma}$ is considered to be 5$\%$ of $I_{\gamma}$.}

	\begin{ruledtabular}
		\begin{tabular}{ccccccccc}
          $E_{\gamma}$(keV)  &  $E_{i}$(keV)  &  $I^{\pi}_{i}$   &   $I^{\pi}_{f}$   &   $I_{\gamma}$   &  $R^{D}_{DCO}$  &  $R^{Q}_{DCO}$ & $P$ &  Multipolarity  \\
          \hline
          288.6(5) & 3880.9(5) &  & 41/2$^{+}$ & 1.7(3) &   &   &   &  \\
          298.0(5) & 2638.0(5) &  & 33/2$^{+}$ & 3.9(5) &   &   &   &  \\
          327.6(1) & 2340.0(1) & 33/2$^{+}$ & 29/2$^{+}$ & 100.0(56) & 1.97(14) & 1.11(8) & 0.33(2) & $E2$ \\
          371.7(1) & 3091.0(5) & 39/2$^{-}$ & 35/2$^{-}$ & 37.0(25) & 2.03(14) & 1.09(8) & 0.27(1) & $E2$ \\
          379.6(1) & 2719.6(1) & 35/2$^{-}$ & 33/2$^{+}$ & 49.0(31) & 0.97(7) & 0.49(3) & 0.14(1) & $E1$ \\
          501.3(1) & 3592.3(1) & 41/2$^{+}$ & 39/2$^{-}$ & 23.4(19) & 1.10(8) & 0.52(4) & 0.48(3) & $E1$ \\
          558.0(3) & 3649.0(3) &  & 39/2$^{-}$ & 3.1(6) &   &   &   &  \\
          751.0(5) & 3091.0(5) & 39/2$^{-}$ & 33/2$^{+}$ & 1.6(3) &  &  &  & $(E3)$\footnote{Based on systematics.} \\
    	\end{tabular}
    \end{ruledtabular}
\end{table*}

Furthermore, some of the inconsistencies in placement of transitions below the 29/2$^{+}$ isomer were resolved in the present work. In the earlier study by Gono \textit{et al.}, a 349-keV transition was placed directly above the 670-keV transition and the 479-keV transition was positioned above the 349-keV transition with an intermediate level at 1019 keV \cite{Gono}. However, this 349-keV transition was later reported to be 348-keV in the study by Decman \textit{et al.} where the ordering of the 479- and 348-keV was interchanged \cite{Decman}. The present investigation confirms the energy of the transition to be 348-keV and the ordering of the 479- and 348-keV transitions as depicted in Fig.~\ref{fig3}. This placement was observed to be in good agreement with that reported by Decman \textit{et al.} \cite{Decman}.

\begin{figure}[b]
	\begin{center}
		\includegraphics[width=1.0\linewidth]{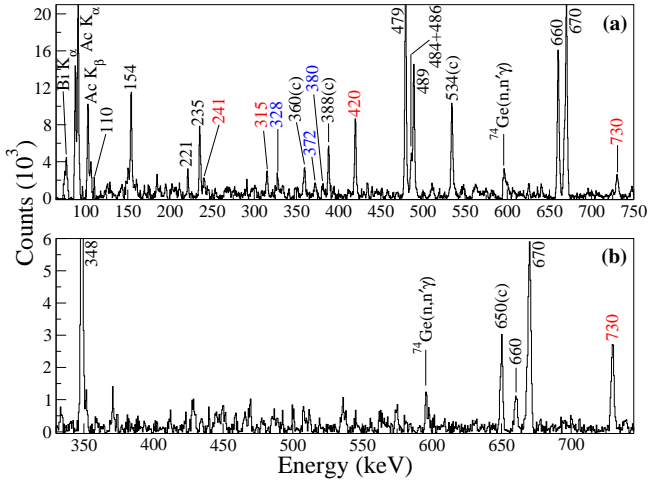}
		\caption{Part of $\gamma$-ray coincidence spectra illustrating the transitions in the gate of the (a) 348-, and (b) 420-keV transitions. The new transitions identified in the present work are shown in red color. The transitions in blue color are those for which the placement was uncertain in the earlier studies and confirmed in the present work. The contaminant transitions are marked with a label (c).}
		\label{fig5}
	\end{center}
\end{figure}

The existence of a few unobserved low-energy transitions in the level scheme was inferred from the coincidence relationships among the observed $\gamma$ rays. For example, an unobserved 10-keV $\gamma$ ray is proposed between the 11/2$^{-}$ and 13/2$^{-}$ states at 670 keV and 660 keV, respectively based on the coincidence relationship between the 660-keV and 479-keV transitions. Also, the 154-keV transition was observed in coincidence with the 348- and 351-keV transitions which suggests the existence of an unobserved 31-keV transition between the levels at 1528 and 1497 keV. The unobserved 31-keV transition was also proposed by Decman \textit{et al.} \cite{Decman,Kondev}.

In addition to the earlier reported transitions, Fig.~\ref{fig3} illustrates around 20 new transitions. A new cascade of two transitions viz. 420- and 730-keV is observed to deexcite the $15/2^{-}$ state to the ground state. The representative coincidence spectra with gates on the 348- and 420-keV transitions are depicted in Fig.~\ref{fig5}. The $I^{\pi}$ = (13/2$^{+}$) may be assigned to the intermediate state at 730 keV based on the sytematics. A similar structure is also known in $^{215}$Fr at 835 keV \cite{KYadav,Decman215Fr,Schulz,Drigert}. An unobserved 60-keV transition is inferred between the (13/2$^{+}$) and $11/2^{-}$ states at 730 keV and 670 keV, respectively,
based on the coincidence spectrum shown in Fig.~\ref{fig5}(b). Furthermore, the coincidence relationship between the 235- and new 315-keV transitions suggests the presence of a 104-keV transition between the levels at 2020 keV and 1916 keV. However, this transition is obscured by the relatively strong Ac x rays.Moreover, several new transitions feeding the known levels are also placed in the level scheme on the basis of coincidence relationships.

\begin{figure}[t]
	\begin{center}
		\includegraphics[width=1.0\linewidth]{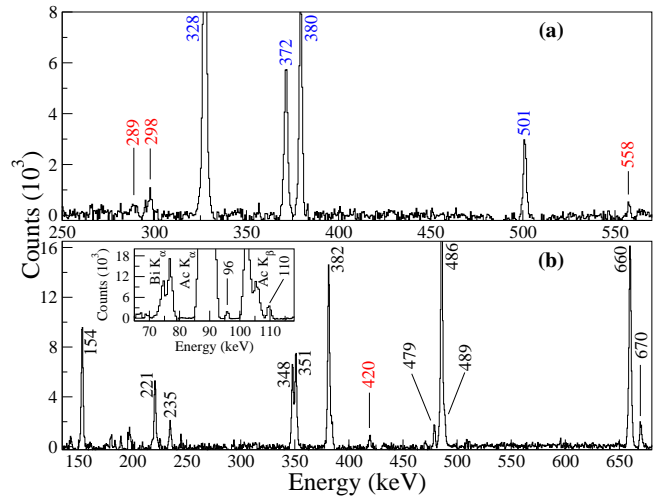}
		\caption{$\gamma$-ray coincidence spectra showing (a) the \textit{early} transitions in the gate of 660-keV $\gamma$ ray, and (b) the \textit{delayed} transitions in the gate of 328-keV $\gamma$ ray within the 200-1200 ns coincidence time window. The new transitions are marked with red color. The transitions whose placement were uncertain in earlier studies and confirmed in the present work are depicted in blue color.}
		\label{fig6}
	\end{center}
\end{figure}

\begin{figure}[t]
	\begin{center}
		\includegraphics[width=1.0\linewidth]{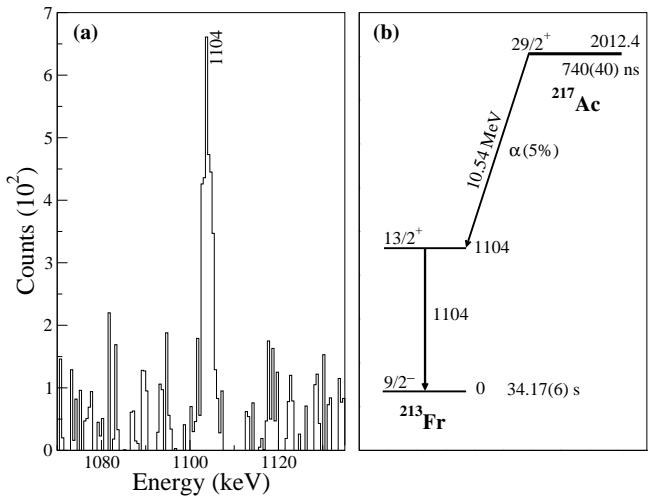}
		\caption{$\alpha$-decay of the 29/2$^{+}$ isomer. Panel (a) illustrates the $\gamma$ ray from $^{213}$Fr in delayed (within 200-1200 ns) coincidence with the 328-keV $\gamma$ ray. A partial decay scheme of the 29/2$^{+}$ isomer is also depicted in panel (b). The data in the panel (b) are taken from the Refs. \cite{Decman,SGarg} and the present work.}
		\label{fig7}
	\end{center}
\end{figure}

The level structure above the 29/2$^{+}$ isomer is observed for the first time as shown in Fig.~\ref{fig3}(B). The \textit{early}-\textit{delayed} $\gamma$-$\gamma$ coincidence technique was employed to identify the transitions above the isomeric state. Figure \ref{fig6} depicts the transitions in \textit{early} and \textit{delayed} coincidence with the 660-keV and 328-keV transitions, respectively, within the 200-1200 ns coincidence time window. In the present study, a total of eight transitions were identified above the isomer, out of which four transitions viz. 328-, 380-, 372-, and 501 keV were already reported in the earlier study by Gono \textit{et al.} \cite{Gono}. However, these transitions and corresponding excited states were not adopted in the recent evaluations due to their tentative nature \cite{Akovali,Kondev}. The placement of these transitions in the level scheme is established on the basis of observed coincidence relationships and intensity considerations. These transitions are marked with blue color in the proposed level scheme as well as in the gated spectra. The $R_{DCO}$ and linear-polarization measurements were carried out to ascertain the spin-parity quantum numbers of the corresponding states at 2340, 2720, 3091 and 3593 keV. These measurements (see Table ~\ref{tab:II}) suggest $E2$ multipolarity for the 328-, and 372-keV $\gamma$ rays and $E1$ multipolarity for the 380- and 501-keV transitions. Thus, $I^{\pi}$ = 33/2$^{+}$, 35/2$^{-}$, 39/2$^{-}$, and 41/2$^{+}$ were inferred for the states at 2340, 2720, 3091 and 3592 keV energy, respectively. In addition, a 751-keV transition was observed in coincidence with the 328- and 501-keV transitions which bypasses the 35/2$^{-}$ state at 2720-keV.
\subsection{Revisiting the $\alpha$-decay of the 29/2$^{+}$ isomer}
The \textit{early}-\textit{delayed} $\gamma$-$\gamma$ coincidence technique was also utilized to revisit the decay of the 29/2$^{+}$ isomer. In addition to the earlier reported transitions in $^{217}$Ac, as illustrated in Fig.~\ref{fig6}(b), a 1104-keV transition of $^{213}$Fr was also observed in the delayed coincidence with the 328-keV $\gamma$ ray [Fig.~\ref{fig7}]. It may be noted that the 1104-keV transition was not observed in the prompt coincidence with any of the known transitions below the 29/2$^{+}$ isomer. Therefore, the possibility that it belongs to $^{217}$Ac may be discarded. Prior to the present work, a 1105-keV transition (i.e. 1104 keV in the present study) in $^{213}$Fr was reported by Decman \textit{et al.,} where it was observed in coincidence with a 10.54-MeV alpha particle depopulating the 29/2$^{+}$ isomer \cite{Decman}. The conversion coefficient of the 1105-keV $\gamma$ ray suggested its $M2$ multipolarity and subsequently the state at 1105 keV was assigned $I^{\pi}$ = 13/2$^{+}$ \cite{Decman}. Furthermore, Decman \textit{et al.} had reported a 498-keV $\gamma$ ray in $^{213}$Fr in coincidence with a 11.13-MeV alpha particle that deexcites the same 29/2$^{+}$ isomer \cite{Decman}. However, the present study doesn't provide any evidence for such a transition. Apart from this, the $\alpha$-decay branch of the 29/2$^{+}$ isomer is determined to be $\sim$ 5$\%$ on the basis of intensity measurements. This is also found to be consistent with that reported in the Ref. \cite{SGarg}.
\subsection{Intensity Measurements}
The relative $\gamma$-ray intensities of the observed transitions were determined using the \textit{early}-\textit{delayed} $\gamma$-$\gamma$ coincidence relationships. The relative intensities of the transitions below the 29/2$^{+}$ isomer were determined using the efficiency corrected \textit{early} gate of the 328-keV transition [Fig.~\ref{fig6}(b)]. However, the weak 730-keV transition was not observed in the delayed coincidence with the 328-keV transition. Therefore, the intensity of the 730-keV transition was obtained in the efficiency corrected prompt gate of the 420-keV transition assuming the total intensity balance at the 730-keV level. In addition, intensities of the remaining new transitions except the 484- and 493-keV transitions were determined in the efficiency corrected prompt gate of the 660-keV transition and normalized with respect to the relative intensity of the suitable transitions. It may be noted that the intensity of the 484-keV transition could not be determined unambiguously in the present work due to the presence of the strong 486-keV transition. Furthermore, intensity of the 493-keV transition was obtained from the efficiency corrected prompt gate of the 486-keV transition and normalized with respect to the relative intensity of the 154-keV transition. Finally, the intensities of all the transitions below the 29/2$^{+}$ isomer were normalized with respect to the 660-keV $\gamma$ ray, intensity of which was assumed to be 100 units. Table ~\ref{tab:I} lists the measured values of the relative intensities of the $\gamma$ rays below the 29/2$^{+}$ isomer. The values are found to be in good agreement with those reported in the Ref. \cite{Decman}.

 The relative intensities of the transitions above and below the 29/2$^{+}$ isomer could not be determined on a single normalizing scale. However, the overall relative variation of $\gamma$ ray intensities of the transitions above the isomer can be determined in the efficiency corrected \textit{delayed} gate of the 660-keV transition. The measured relative intensities are listed in Table ~\ref{tab:II}. Figure~\ref{fig6}(a) depicts the \textit{early} transitions in the gate of the 660-keV $\gamma$ ray. The weak 751-keV $\gamma$ ray was not observed in this gate. Thus, the intensity of this transition was determined in the efficiency corrected prompt gate of the 328-keV $\gamma$ ray and normalized with respect to the relative intensity of the 501-keV transition. Finally, the relative intensities of all the transitions above the 29/2$^{+}$ isomer are normalized with respect to the 328-keV $\gamma$ ray by assuming its relative intensity to be 100 units. One of the intriguing aspects of the present work is the missing intensity at the 2340 keV level. Various possibilities such as isomeric nature of the 33/2$^{+}$ state and presence of unobserved low-energy highly-converted transitions feeding the 33/2$^{+}$ state, were explored to understand this. However, the experimental data do not allow to draw any meaningful understanding of the missing intensity.
\section{Discussion} \label{section3}
 Excited states in nuclei near the doubly-magic $^{208}$Pb are well known examples of underlying spherical nuclear shell structure. For example, the level structure of the closed shell $N= 126$ isotones (e.g. $^{211}$At, $^{213}$Fr and $^{215}$Ac) have been accounted well by the shell-model configurations \cite{Bayer,Byrne3, Decman215Ac}. Similarly, the $N=128$ isotones (e.g. $^{213}$At and $^{215}$Fr) exhibit single particle structures because of their proximity to the shell closures. Therefore, the excited states in such nuclei can be interpreted as coupling of odd proton to the even-even core states  \cite{Sjoreen, Lane,Decman215Fr,Drigert,KYadav}. 
 
\begin{figure}[b]
 	\begin{center}
 		\includegraphics[width=1.0\linewidth]{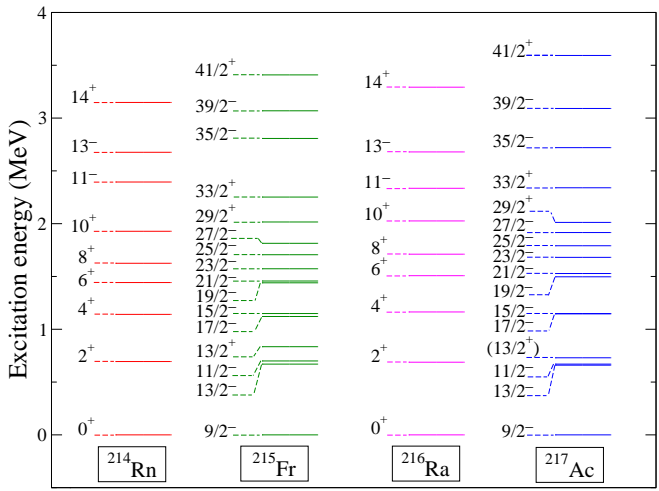}
 		\caption{Level energy systematics of the $N=128$ isotones. The data are taken from Refs. \cite{Dracoulis,KYadav, Muralithar} and the present work.}
 		\label{fig8}
 	\end{center}
 \end{figure}

 $^{217}$Ac has seven valence protons and two valence neutrons outside the $^{208}$Pb core. Therefore, it is expected to exhibit single-particle structures. However, it is also known that addition of an extra neutron to $^{217}$Ac leads to the octupole-correlation in $^{218}$Ac \cite{Decman,Debray}. In the earlier studies, the reported level structure of $^{217}$Ac was described in terms of the spherical shell model and deformed independent-particle model \cite{Decman,Decman215Fr}. From the comparison of level structures in the neighboring $N=128$ isotones, it is observed that the excited states in $^{217}$Ac resemble those in $^{216}$Ra as illustrated in Fig.~\ref{fig8}. The excitation energies of the levels remain almost constant irrespective of the variation in proton numbers, which indicates that the neutron excitations might play an important role in governing the structures in these nuclei \cite{Decman}. Therefore, a detailed shell-model calculation is required to elucidate the level structures of $^{217}$Ac.
 
A large-scale shell-model calculations have been performed to compare the experimental observations with the theoretical predictions. The $82 \le Z \le  126$ and $126 \le N \le 184$ model space has been utilized for the calculations, where protons occupy the $1h_{9/2}$, $2f_{7/2}$, $2f_{5/2}$, $3p_{3/2}$, $3p_{1/2}$, and $1i_{13/2}$ orbitals and neutrons occupy the $1i_{11/2}$, $2g_{9/2}$, $2g_{7/2}$, $3d_{5/2}$, $3d_{3/2}$, $4s_{1/2}$, and $1j_{15/2}$ orbitals. In this model space, the Kuo-Herling particle effective KHPE interaction from Ref. \cite{KHPE} has been employed. This effective residual interaction was originally derived from a free \textit{NN} potential of Hamada and Johnston using a renormalization technique developed by Kuo and Brown \cite{KHPE}. The model space is not truncated for the present calculations, allowing the valence nucleons to occupy the available orbitals independently. In order to diagonalize the Hamiltonian, a parallel shell-model code KSHELL has been utilized \cite{KShell}. It is noted that the shell-model calculations for $^{215}$Ac using the same interaction was reported in the  Ref.~\cite{Bharti} where the results were seen to agree well with the experimental observations. 

\begin{figure}[t]
	\begin{center}
		\includegraphics[width=1.0\linewidth]{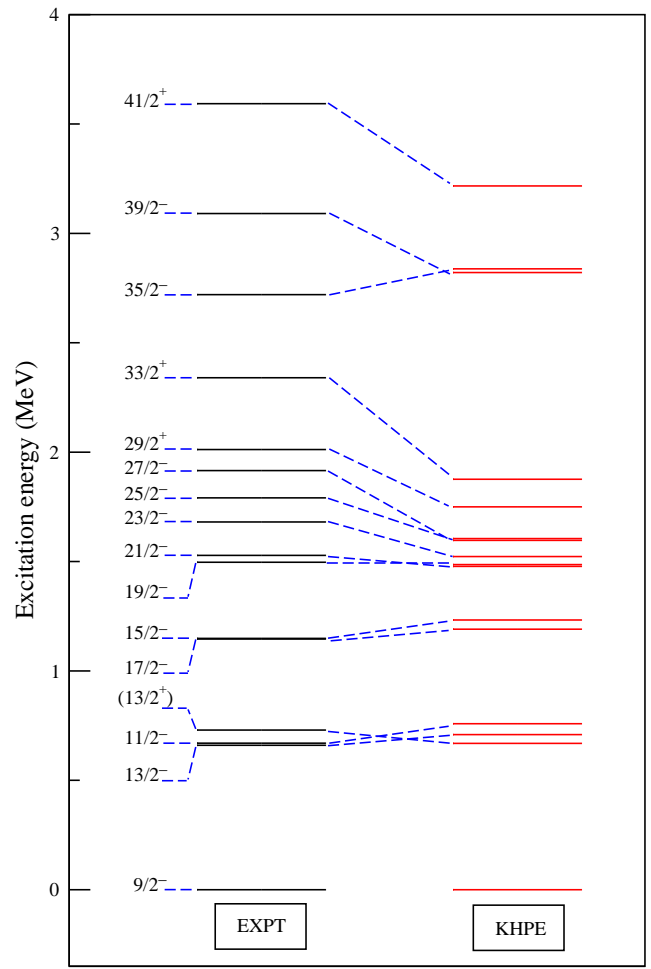}
		\caption{A comparison of the experimental results in $^{217}$Ac with the shell-model calculations using the KHPE interaction. Only yrast and near-yrast states with firm spin-parity assignments are considered for the comparison.}
		\label{fig9}
	\end{center}
\end{figure}

A comparison of the experimental results and the shell-model predictions for $^{217}$Ac is illustrated in Fig.~\ref{fig9}. The observed level energies for the yrast and near-yrast states below the $29/2^{+}$ are found to be in good agreement (to within 300 keV) with those predicted by the shell model. The low-lying negative-parity states up to the $25/2^{-}$ have almost similar contributions from the $\pi(h^{5}_{9/2} f^{2}_{7/2})\otimes\nu(g^{2}_{9/2})$, $\pi(h^{3}_{9/2} f^{2}_{7/2} i^{2}_{13/2})\otimes\nu(g^{2}_{9/2})$, and $\pi(h^{5}_{9/2} i^{2}_{13/2})\otimes\nu(g^{2}_{9/2})$ configurations. This suggests a significant configuration mixing (even for the ground state), which is considered as an indication of enhanced p-n interaction for the underlying orbitals \cite{Decman215Fr}. The \textit{g}-factor measurement for the ground state by Decman \textit{et al.} also supports the mixing \cite{Decman}. The dominant configurations for the $27/2^{-}$ state are predicted to be $\pi(h^{5}_{9/2} i^{2}_{13/2})\otimes\nu(i^{1}_{11/2} g^{1}_{9/2})$, $\pi(h^{5}_{9/2} f^{2}_{7/2})\otimes\nu(i^{1}_{11/2} g^{1}_{9/2})$, and $\pi(h^{3}_{9/2} f^{2}_{7/2} i^{2}_{13/2})\otimes\nu(i^{1}_{11/2} g^{1}_{9/2})$, with probabilities over $17-19\%$. The $29/2^{+}$ isomeric state appears to be mixed with major contributions ($17-23\%$) from the $\pi(h^{4}_{9/2} f^{2}_{7/2} i^{1}_{13/2})\otimes\nu(g^{2}_{9/2})$, $\pi(h^{6}_{9/2} i^{1}_{13/2})\otimes\nu(g^{2}_{9/2})$, and $\pi(h^{4}_{9/2} i^{3}_{13/2})\otimes\nu(g^{2}_{9/2})$ configurations. It may be noted that the $29/2^{+}$ isomeric state decays to the $27/2^{-}$ and $25/2^{-}$ states. The shell-model configurations suggest a change in the occupied proton and neutron orbitals for these states. Consequently, this configuration change hinders the decay of the $29/2^{+}$ state.
 
Furthermore, the newly established levels above the $29/2^{+}$ isomer are also compared with the shell-model predictions. The primary configuration for the $33/2^{+}$ and $41/2^{+}$ states is predicted to be $\pi(h^{4}_{9/2} f^{2}_{7/2} i^{1}_{13/2})\otimes\nu(i^{1}_{11/2} g^{1}_{9/2})$. The dominant configuration for the $35/2^{-}$ state is $\pi(h^{6}_{9/2} i^{1}_{13/2})\otimes\nu(g^{1}_{9/2} j^{1}_{15/2})$ whereas the $\pi(h^{6}_{9/2} i^{1}_{13/2})\otimes\nu(i^{1}_{11/2} j^{1}_{15/2})$ configuration is suggested for the $39/2^{-}$ state. In addition, the presence of a second dominant configuration [$\pi(h^{4}_{9/2} f^{2}_{7/2} i^{1}_{13/2})\otimes\nu(i^{1}_{11/2} j^{1}_{15/2})$] for the $39/2^{-}$ state with almost equal parentage ($\sim23\%$) allows the $(\nu j_{15/2}\rightarrow g_{9/2})$ $E3$ transition to the $33/2^{+}$ state, similar to the case of $^{215}$Fr \cite{KYadav, Decman215Fr}. Moreover, the presence of the $i_{13/2}$ orbital in the proton configurations indicates its importance in the structure of the yrast states above the isomer. It may be noted that the shell-model calculations are not able to account well for the states above the $29/2^{+}$ isomer. This may be ascribed to the effects which could not taken into account by the shell model.

The yrast and near-yrast states in $^{217}$Ac may also be understood using a semi-empirical approach based on the energy-level systematics of the $N=128$ isotones as shown in Fig.~\ref{fig8}. The low-lying states in $^{217}$Ac are compared with those of $^{216}$Ra. The observed similarity in the excitation energies suggests that the yrast states in $^{217}$Ac may be interpreted in terms of the coupling of the odd $h_{9/2}$ proton to the even-even $^{216}$Ra core. For example, the multiplets with spin-parity (11/2$^{-}$, 13/2$^{-}$), (15/2$^{-}$, 17/2$^{-}$), (19/2$^{-}$, 21/2$^{-}$) and (23/2$^{-}$, 25/2$^{-}$) can be realized through the $\pi (h_{9/2}) \otimes [(2^{+}-8^{+})^{216}$Ra] configuration \cite{Decman, Muralithar}. A similar structure was also observed in the low-lying states of $^{215}$Fr \cite{KYadav}. In addition, a (13/2$^{+}$) state is identified based on the systematics. This 13/2$^{+}$ state in $^{215}$Fr was interpreted as a coupling of the proton $i_{13/2}$ level to the ground state of the $^{214}$Rn core \cite{KYadav}. Similarly, the (13/2$^{+}$) state in $^{217}$Ac may also be understood in terms of the $\pi (i_{13/2}) \otimes [0^{+}(^{216}$Ra)] configuration, which is also found to be consistent with the one predicted by the shell-model [$\pi(h^{4}_{9/2} f^{2}_{7/2} i^{1}_{13/2})\otimes\nu(g^{2}_{9/2})$]. Furthermore, the $29/2^{+}$ isomeric state may be realized from the coupling of the $i_{13/2}$ proton to the $8^{+}$ state of $^{216}$Ra \cite{Decman}. This interpretation is further corroborated by the shell-model predictions as well as alpha decay of the $29/2^{+}$ isomer, which is discussed later.
 
Similar to the low-lying states, the yrast structures above the $29/2^{+}$ isomer may be compared to the high-spin states of $^{216}$Ra. The observed similarity in the level structures suggests that the $33/2^{+}$, $35/2^{-}$, and $39/2^{-}$ states in $^{217}$Ac may be understood as arising from the $\pi i_{13/2} \otimes [10^{+}, 11^{-},$ and $13^{-} (^{216}$Ra)] configurations, respectively. These configurations are found to be in good agreement with the shell-model calculations. In addition, the above interpretation is also supported by the systematics of the other $N=128$ isotones. For example, a striking resemblance is observed in the level structures of $^{217}$Ac and $^{215}$Fr. The states above the $29/2^{+}$ isomer in $^{217}$Ac may also be understood in a similar manner to those in $^{215}$Fr. In $^{215}$Fr, a $39/2^{-}$ isomeric state with $T_{1/2} = 11.4(14) $ ns at 3208 keV decays to another isomeric $33/2^{+}$ state [$T_{1/2} = 7.1(15)$ ns] at 2252 keV via two decay paths \cite{KYadav}. One of the decay paths proceeds via a 262-keV $E2$ transition followed by a 555-keV $E1$ transition, whereas the other decay path involves a direct 817-keV $E3$ transition to the $33/2^{+}$ state. A similar decay pattern is also observed in the present case for the $39/2^{-}$ state at 3091 keV. This state deexcites to the $33/2^{+}$ state via the 372-keV $E2$ transition followed by the 380-keV $E1$ transition. Alternatively, the 751-keV transition has been identified which directly depopulates the $39/2^{-}$ state to the $33/2^{+}$ state. The observed similarity in the level structure of $^{217}$Ac and $^{215}$Fr also supports the $E3$ multipolarity of this transition. Furthermore, the yrast states above the $29/2^{+}$ isomer in $^{215}$Fr were interpreted in terms of the coupling of the $i_{13/2}$ proton to the $10^{+}, 11^{-},$ and $13^{-}$ states of $^{214}$Rn core \cite{KYadav}. Therefore, a similar interpretation may also be valid for $^{217}$Ac with the $^{216}$Ra as a core.
 
The systematics of nuclei with $N=128$ neutrons illustrates a pronounced alpha-decay widths in their ground states as well as in some excited states \cite{Decman, Nomura}.  This may be attributed to the extra stability gain in the daughter nuclides because of the $N=126$ shell closure. The $Q_{\alpha}$ value becomes maximum for the $N=128$ isotones \cite{Nomura}. The alpha decay in the $N=128$ isotones has been of considerable interest because of the exceptionally short lifetime \cite{Decman}. It may be noted that the $^{217}$Ac is well known for its fastest alpha decay in the ground state \cite{Decman}. In addition to the ground state, several excited states were reported to decay via alpha emission in the previous studies \cite{Decman, Gono}. However, alpha decay of only $29/2^{+}$ state could be revisited. This decay may be understood in parallel to that of the $8^{+}$ isomeric state in $^{216}$Ra, which decays to the ground state of $^{212}$Rn via alpha emission \cite{Dracoulis216Ra,Decman}. A similar decay pattern is also observed in $^{217}$Ac where the $29/2^{+}$ isomer decays to the $13/2^{+}$ state of $^{213}$Fr \cite{Decman}. The observed similarity suggests that the $29/2^{+}$ isomer in $^{217}$Ac may be understood as resulting from the coupling of an extra $i_{13/2}$ proton to the $8^{+}$ state in $^{216}$Ra i.e. [$\pi i_{13/2}$ $\otimes$ $8^{+}(^{216}$Ra)]$_{29/2^{+}}$ \cite{Decman}. Similarly, the $13/2^{+}$ state in $^{213}$Fr may be interpreted as the coupling of an $i_{13/2}$ proton to the ground state of $^{212}$Rn i.e. [$\pi i_{13/2}$ $\otimes$ $0^{+}(^{212}$Rn)]$_{13/2^{+}}$ \cite{Decman}. The empirical configuration for the $13/2^{+}$ state in $^{213}$Fr agrees well with the shell-model predictions which is dominantly $\pi(h^{4}_{9/2} i^{1}_{13/2})$. In the other words, the shell-model configurations suggest a considerable overlap in the wave functions of the $29/2^{+}$ state in $^{217}$Ac and the $13/2^{+}$ state in $^{213}$Fr which favors the alpha decay in $^{217}$Ac, despite the angular momentum change of $8\hbar$. Therefore, both the shell-model and semi-empirical interpretations delineate the importance of the $i_{13/2}$ proton orbital in the alpha decay of the $29/2^{+}$ isomer.
\section{summary}
Excited states in the transitional $^{217}$Ac nucleus have been studied using in-beam $\gamma$-ray spectroscopic techniques following the $^{209}$Bi($^{12}$C,4n)$^{217}$Ac reaction. The level structure has been extended up to 3.8 MeV excitation energy and $I^{\pi} = 41/2^{+}$, with the addition of around 20 new transitions. The yrast and near-yrast states below the $29/2^{+}$ isomer have been revisited. Some of the inconsistencies in the earlier level schemes of $^{217}$Ac have been resolved. The present investigation establishes the level structure above the $29/2^{+}$ isomer for the first time.  Large-scale shell-model calculations have been performed using the KHPE interaction. The calculations account well for the states below the $29/2^{+}$ isomer, while considerable deviations are observed for the states above the isomer. The excited states in $^{217}$Ac have been compared with the states in the other $N = 128$ isotones, which provides further insight into the level structure of $^{217}$Ac. The yrast and near-yrast states below the $29/2^{+}$ isomer may be understood in terms of the coupling of an odd $h_{9/2}$ proton to the states in the even-even $^{216}$Ra core. Conversely, the states above the $29/2^{+}$ isomer are attributed to the coupling of the odd $i_{13/2}$ proton to the states in the $^{216}$Ra core. These interpretations align well with the shell-model configurations.  Furthermore, the present study revisits the alpha decay of the $29/2^{+}$ isomer in $^{217}$Ac and the overall decay scheme has been interpreted in the framework of shell model. The predicted shell-model configurations highlights the importance of the $i_{13/2}$ proton orbital in the alpha decay of the $29/2^{+}$ state.
\section{Acknowledgment}
The authors would like to thank the staff of TIFR Target Laboratory for their support in target preparation. The technical assistance from INGA and PLF staffs of TIFR during the course of experiment is highly appreciated. DS and AYD would like to acknowledge the financial support by DST-SERB vide grant no. CRG/2020/002169. Madhu acknowledges the financial support from DST, India under the INSPIRE fellowship scheme (IF 180082). This work is supported by the Department of Atomic Energy, Government of India (Project Identification No. RTI 4002), and the Department of Science and Technology, Government of India (Grant No. IR/S2/PF-03/2003-II). AK acknowledges the support of the Program for promoting research on the supercomputer Fugaku, MEXT, Japan (JPMXP1020230411).

\end{document}